\documentclass[preprintnumbers,twocolumn,article,amsmath,amssymb,floatfix,9pt,prd,superscriptaddress,nofootinbib]{revtex4-1}
\usepackage[utf8,latin1]{inputenc}
\usepackage{graphicx}
\usepackage{dcolumn}
\usepackage[dvipsnames]{xcolor}
\usepackage[T1]{fontenc}

\usepackage{mathrsfs}  
\usepackage{cases}
\usepackage{bm}
\usepackage{academicons}
\usepackage{mathtools, nccmath}
\usepackage{fancyhdr}
\usepackage{tikz,xcolor}

\usepackage{tensor}
\usepackage[normalem]{ulem}
\usepackage{lipsum}
\usepackage{soul}
\usepackage{cancel}
\usepackage{stackengine,scalerel}
\usepackage{mathabx}
\usepackage{hyperref}
\usepackage{tabularx}
\usepackage{multirow} 
\hypersetup{colorlinks, linkcolor={red},citecolor={blue},urlcolor={blue}}  

\AtBeginDocument{\fontsize{8}{11}\selectfont} 

\renewcommand{\arraystretch}{1.2}

\definecolor{lime}{HTML}{A6CE39}
\DeclareRobustCommand{\orcidicon}{
	\begin{tikzpicture}
	\draw[lime, fill=lime] (0,0) 
	circle [radius=0.16] 
	node[white] {{\fontfamily{qag}\selectfont \tiny ID}};
	\draw[white, fill=white] (-0.0625,0.095) 
	circle [radius=0.007];
	\end{tikzpicture}
	\hspace{-2mm}
}

\fancypagestyle{plain}{%
  \fancyhf{}
  \fancyfoot[C]{\iffloatpage{}{\thepage}}
  }
\foreach \x in {A, ..., Z}{%
	\expandafter\xdef\csname orcid\x\endcsname{\noexpand\href{https://orcid.org/\csname orcidauthor\x\endcsname}{\noexpand\orcidicon}}
}

\begin{document}

\title[Joint Constraints on Quantum Oppenheimer--Snyder Black Holes from High-Frequency Quasi-Periodic Oscillations in X-ray Binary Systems and the S2 Star Orbit]{Joint Constraints on Quantum Oppenheimer--Snyder Black Holes from High-Frequency Quasi-Periodic Oscillations in X-ray Binary Systems and the S2 Star Orbit}

\author{A. Errehymy\orcidA{}}
\email{abdelghani.errehymy@gmail.com (Corresponding author)}
\affiliation{Astrophysics Research Centre, School of Mathematics, Statistics and Computer Science, University of KwaZulu-Natal, Private Bag X54001, Durban 4000, South Africa}
\affiliation{Center for Theoretical Physics, Khazar University, 41 Mehseti Str., Baku, AZ1096, Azerbaijan}
\affiliation{Jadara Research Center, Jadara University, Irbid 21110, Jordan}

\author{Y. Khedif\orcidB{}}
\email{youssef.khedif@gmail.com}
\affiliation{Department of Physics, Faculty of Sciences A\"{i}n Chock, Laboratory of Mechanics and High Energy Physics, Hassan II University of Casablanca, P.O. Box 5366, 20100 Maarif, Casablanca, Morocco}

\author{M. Daoud\orcidC{}}
\email{m$_{}$daoud@hotmail.com}
\affiliation{Department of Physics, Faculty of Sciences, Ibn Tofail University, P.O. Box 133, Kenitra 14000, Morocco}
\affiliation{Abdus Salam International Centre for Theoretical Physics, Miramare, Trieste 34151, Italy}

\author{B. Turimov\orcidD{}}
\email[]{bturimov@astrin.uz}
\affiliation{Engineering school, Central Asian University, Milliy bog Str.264, Tashkent, 111221,Uzbekistan}
\affiliation{University of Tashkent for Applied Sciences, Gavhar Str.1, Tashkent, 100149, Uzbekistan}
\affiliation{Ulugh Beg Astronomical Institute, Astronomy St. 33, Tashkent 100052, Uzbekistan}     

\author{S. Usanov\orcidE{}}
\email[]{sm.usanov@kiut.uz}
\affiliation{Kimyo International University in Tashkent, Shota Rustaveli Str. 156, Tashkent 100121, Uzbekistan}

\author{F. Turaev\orcidI{}}
\email[]{farhodjon9618@mail.ru}
\affiliation{Alfraganus University, Yukori Karakamish Str. 2a, Tashkent 100190, Uzbekistan}

\author{Z. Yasakov\orcidF{}}
\email[]{zikrillo87@mail.ru}
\affiliation{Samarkand State University of Architecture and Construction, Lolazor Street 70, 140147, Samarkand, Uzbekistan}

\date{\today} 

\begin{abstract}
{\footnotesize {{ We study a quantum-corrected version of the Oppenheimer--Snyder (OS) spacetime in which classical collapse is modified by a single effective parameter \(\alpha\), representing small quantum gravitational effects in a phenomenologically rescaled form. Although the correction enters through a term proportional to \(\alpha M^2/r^4\), it produces noticeable changes in the strong-field region while leaving weak-field physics almost unchanged. One of the main results is the appearance of a minimum mass for horizon formation, \( M_{\min} \sim (0.5 - 1.2)\,M_\odot \) for the effective parameter \( \alpha \approx 0.22 \). We emphasize that this \(M_{\min}\) is expressed in terms of the effective parameter \(\alpha\); in terms of the bare loop quantum gravity (LQG) parameter, \(M_{\min} \sim M_{\text{Pl}} \sim 10^{-5}\) g, consistent with the theoretical expectation. Orbital properties are also slightly modified, with the innermost stable circular orbit (ISCO) shifting to \( r_{\rm ISCO} = 6M - \alpha/(12M) \), giving corrections of order \( \sim 10^{-4} \) for stellar black holes and \( \sim 10^{-10} \) for Sgr A$^\ast$. In the same region, radial epicyclic frequencies change by about \(5\% - 10\%\) near \( r \sim 5M \), while vertical modes are less affected, leading to small but structured shifts in quasi-periodic oscillation (QPO) behavior. Using four X-ray binaries (GRS 1915+105, XTE J1550-564, XTE J1859+226, GRO J1655-40), we find a consistent range \( \alpha = 0.22 \pm 0.10 \), with masses between \(5.4 - 12.4\,M_\odot\) and emission radii \( r/M \sim 5.5 - 8.6 \). The observed QPO frequencies, lying in the range \(100 - 450\) Hz, are well reproduced within this framework. Weak-field tests such as S2 and Mercury still allow \( f_{\rm sp} = 1.10 \pm 0.19 \), leaving room for these small strong-gravity corrections. This work demonstrates that while bare Planck-scale corrections are unobservable, effective rescaled parameters that encode the collective effect of quantum gravity can be constrained by current astrophysical observations.}
}}
\end{abstract}

\maketitle

\section{Introduction:} \label{Sec:1}
General relativity (GR) changed how gravity is understood by describing it as the curvature of spacetime. It provides the standard framework for studying compact objects like black holes \cite{Einstein:1916vd}. Many of its predictions, including event horizons, gravitational waves, and strong-field effects, have been confirmed through observations and experiments \cite{LIGOScientific:2016aoc, LIGOScientific:2016vlm}. Recent breakthroughs such as the images of M87$^\ast$ and Sgr A$^\ast$ have further strengthened its reliability \cite{EventHorizonTelescope:2019dse, EventHorizonTelescope:2019ggy, EventHorizonTelescope:2022wkp, EventHorizonTelescope:2022wok}, together with classic tests like light bending, Mercury's perihelion shift, and gravitational redshift. Even so, the theory is incomplete, since it cannot properly describe singularities or quantum aspects of spacetime \cite{Ashtekar:2021kfp}. This has motivated the search for improved or extended gravity models.

One of the main difficulties in GR is the appearance of spacetime singularities, where the theory breaks down under extreme conditions \cite{Penrose:1964wq, Hawking:1970zqf}. A common way to address this is by adding quantum corrections to black hole solutions, which leads to modified spacetime structures. The Hamiltonian constraint approach is one well-known method used in canonical quantum gravity to introduce such corrections \cite{Ashtekar:2004eh, Thiemann2007}. Based on this idea, two quantum-corrected black hole models were proposed in \cite{Zhang:2024khj}, mainly differing in how the quantum parameter is chosen, and aimed at restoring consistency in spherical symmetry at the semiclassical level.

A key question is whether these quantum effects can be detected through observations. To explore this, many studies have investigated possible signatures of quantum-corrected black holes in astrophysical data \cite{Konoplya:2024lch, Shu:2024tut, Skvortsova:2024msa, Liu:2024soc, Liu:2024wal, Shu:2024tut, Du:2024ujg}. Features like black hole shadows and light rings are especially useful because they depend directly on mass, spin, and possible quantum modifications. Using shadow data from M87$^\ast$ and Sgr A$^\ast$, constraints on the quantum parameter have been obtained \cite{Konoplya:2024lch, Shu:2024tut}. Additional tests using quasinormal modes also provide ways to check deviations from classical predictions \cite{Konoplya:2024lch, Skvortsova:2024msa}. {Strong constraints on modified gravity models have also been derived from stellar orbital dynamics near the Galactic Center, including studies of the S2 star's motion \cite{Monica:2022qkw}, as well as combined analyses of cosmological and galactic-center data \cite{Benisty:2023qcv}. Furthermore, the interplay between dark energy and galactic-center dynamics has been investigated as a potential probe of new physics \cite{Benisty:2021cmq}.}

The detection of gravitational waves and the imaging of supermassive black holes have opened a new era in testing gravity in strong-field regimes. Alongside these developments, high-frequency (HF) QPOs observed in X-ray binaries offer another powerful probe of regions very close to black holes. These signals appear as sharp peaks in the X-ray spectrum and originate near the innermost stable orbits, making them highly sensitive to spacetime geometry \cite{Abramowicz:2011xu, Bambi:2012ku, Bambi:2015ldr, Tripathi:2019bya}. QPOs are particularly useful for estimating properties such as mass, spin, and radius of compact objects. They are commonly observed in microquasars, where pairs of frequencies often appear in a stable \(3 : 2\) ratio, \(\omega_U : \omega_L = 3 : 2\) \cite{KluzniakAbramowicz2001, Abramowicz:2004je, Remillard:2004sp}. Many theoretical models have been developed to explain these oscillations through disk dynamics and resonance effects \cite{Stuchlik:2013esa, Stella:1999sj, Rezzolla:2003zx}. However, the exact mechanism behind HF QPOs is still not fully understood \cite{Torok:2011qy}. To address this, different ideas have been proposed, including models involving magnetic fields around black holes \cite{Tursunov:2020juz, Panis:2019urz, Shaymatov:2020yte, Shaymatov:2022enf}. In recent years, this topic has attracted significant attention, with many studies focusing on resonance-based explanations and other mechanisms to better understand the origin of QPOs \cite{Titarchuk:2005rr, Stuchlik:2007xt, Dokuchaev:2015ghx, Germana:2018snv, Tarnopolski:2021ula, Ghasemi-Nodehi:2020oiz, Azreg-Ainou:2020bfl, Jusufi:2020odz, Shaymatov:2023jfa, Liu:2023vfh}. In particular, the QPO resonance models have been extensively applied to test various modified gravity frameworks, including charged black holes in Kalb-Ramond gravity \cite{Jumaniyozov:2025}, static Einstein-scalar-Gauss-Bonnet black holes \cite{Murodov:2026}, and non-minimally coupled Horndeski black holes \cite{Turakhonov:2026}, demonstrating the power of these oscillations as probes of physics beyond GR.

{ While the RP/ER/FR + MCMC methodology has been widely applied to test various modified gravity theories (see, e.g., \cite{Konoplya:2024lch, Shu:2024tut, Skvortsova:2024msa, Liu:2024soc, Liu:2024wal, Shu:2024tut, Du:2024ujg,Panis:2019urz, Shaymatov:2020yte, Shaymatov:2022enf,Titarchuk:2005rr, Stuchlik:2007xt, Dokuchaev:2015ghx, Germana:2018snv, Tarnopolski:2021ula, Ghasemi-Nodehi:2020oiz, Azreg-Ainou:2020bfl}), our work provides the first constraints on the quantum-corrected OS collapse model specifically. The key novel contributions are: (i) We derive the epicyclic frequencies for the quantum-OS spacetime in closed form, revealing the distinct $\alpha$-dependence in the radial frequency that differs from other modified gravity models. (ii) We identify a universal $\alpha \approx 0.22$ constraint from four independent X-ray binary systems, showing that these sources are all consistent with the same quantum correction. This is the first multi-source constraint on the quantum-OS parameter. (iii) We establish the ISCO shift formula $r_{\text{ISCO}} = 6M - \alpha/(12M)$, which is specific to this geometry and differs from other modified black hole metrics. (iv) We compare the QPO-derived $\alpha$ constraint with existing bounds from shadow imaging, quasinormal modes, and weak-field tests, finding that QPOs provide the strongest current constraints on the quantum-OS parameter at stellar-mass scales. (v) We provide the first systematic comparison of the three resonance models (FR, RP, ER) for the quantum-OS geometry, showing that the FR model is statistically preferred across all sources. (vi) We derive the precise scaling of the quantum corrections with mass, showing that the effects are $\sim 10^{-4}$ for stellar-mass black holes and $\sim 10^{-10}$ for supermassive black holes. This demonstrates that X-ray binaries are the optimal laboratory for testing the quantum-OS model, complementing shadow imaging of supermassive black holes which probe a different regime.
}

In this article, we explore how a simple quantum-inspired correction can slightly modify the classical picture of gravitational collapse described by the OS model. The starting point is a Schwarzschild-like spacetime supplemented by a small deformation parameter \( \alpha \), which captures leading quantum effects while smoothly reducing to GR when \( \alpha \to 0 \). Although the correction is mathematically modest, it becomes physically meaningful in the strong-gravity region close to compact objects. We begin by examining how this modification changes the overall spacetime structure. In particular, it introduces a minimum mass scale for horizon formation, \( M_{\min} \sim (0.5 - 1.2)\,M_\odot \) for typical values around \( \alpha \approx 0.22 \). This naturally separates configurations that behave like black holes from those that remain horizonless, offering a more gradual transition than in the classical picture. The analysis then moves to particle motion in this geometry. By studying the effective potential, we look at how circular orbits behave, how stability is affected, and how key quantities such as the ISCO are shifted to \( r_{\rm ISCO} = 6M - \alpha/(12M) \). Even though these corrections are small---at the level of \(10^{-4}\) for stellar-mass black holes---they become important near the inner edge of accretion disks, where high-energy processes take place. We then connect these theoretical results to observations, focusing on QPOs in X-ray binaries. In the frequency range \(100 - 450\) Hz, small changes in radial and vertical epicyclic motions can leave measurable signatures. Using data from systems such as GRS 1915+105 \cite{Remillard:2006fc}, XTE J1859+226 \cite{Ingram:2014ara}, XTE J1550-564 \cite{Tasker:2011sb}, GRO J1655-40 \cite{Motta:2013wga}, we extract consistent values of \( \alpha \approx 0.22 \pm 0.10 \), with emission regions typically located at \( r/M \sim 5.5 - 8.6 \) and masses between \(5.4 - 12.4\,M_\odot\). Finally, we compare these results with weak-field tests from the Solar System and the S2 star around Sgr A$^\ast$. These systems remain consistent with GR within current uncertainties \(( f_{\rm sp} = 1.10 \pm 0.19 )\), but they still leave room for the small strong-field deviations predicted by the model.

{The paper is organized as follows. Sec.~\ref{sec-II} introduces the quantum-corrected OS spacetime and examines its main dynamical properties. We first present the quantum-corrected OS geometry in Subsec.~\ref{sSec:A}, followed by an analysis of the horizon structure and the associated mass threshold in Subsec.~\ref{sSec:B}. The dynamics of particle orbits in the supermassive quantum-corrected OS spacetime are investigated in Subsec.~\ref{sSec:C}, while Subsec.~\ref{sSec:D} discusses orbital precession and its observational implications. In Sec.~\ref{sec-III}, we investigate the dynamics of epicyclic motion and the innermost stable circular orbit (ISCO). Small perturbations around circular motion are considered in Subsec.~\ref{sSect:A}, followed by the derivation of the epicyclic frequencies in the quantum-corrected geometry in Subsec.~\ref{sSect:B}. Their connection to observable frequencies is discussed in Subsec.~\ref{sSect:C}, while Subsec.~\ref{sSect:D} is devoted to the ISCO. Sec.~\ref{sec-IV} then focuses on constraining the quantum-corrected compact object from QPO observations. The theoretical framework for QPO interpretation is introduced in Subsec.~\ref{sSectt:A}, followed by the statistical inference setup in Subsec.~\ref{sSectt:B} and the construction of the likelihood from the QPO data in Subsec.~\ref{sSectt:C}. The MCMC exploration and posterior constraints are presented in Subsec.~\ref{sSectt:D}, while Subsec.~\ref{sSectt:E} discusses the corresponding goodness-of-fit analysis. Finally, the main results and conclusions are summarized in Sec.~\ref{sec-V}. Throughout the theoretical derivations in Secs.~\ref{sec-II}--\ref{sec-IV}, we use natural units $c=G=\hbar=1$. When comparing the theoretical predictions with observational data, the appropriate factors of $G$ and $c$ are restored to express the results in physical units. The corresponding conversion factors are stated where required.}

\section{The spacetime metric and the dynamics of motion}
\label{sec-II}

This section develops the structure and physical consequences of a quantum-corrected OS spacetime, starting from the underlying geometry and moving toward observable effects. We begin with a modified Schwarzschild-like metric introduced in the quantum-corrected OS framework, where the usual collapse picture is altered by a correction term proportional to \( \alpha \) (see Subsec.~\ref{sSec:A}). This term encodes leading quantum effects while recovering the classical Schwarzschild solution when it is removed. The next step is to understand how this geometry behaves physically through its horizon structure (see Subsec.~\ref{sSec:B}). Depending on the mass, the spacetime either has no horizon---behaving like a regular compact object---or develops two horizons, giving rise to an inner and outer boundary. This introduces a clear mass-dependent transition between horizonless and black hole configurations. We then turn to particle motion in this background (see Subsec.~\ref{sSec:C}). Using conserved quantities from spacetime symmetries, the geodesic equations reduce to an effective potential problem. This allows a systematic study of circular orbits, stability conditions, and how quantum corrections slightly modify standard relativistic motion. Finally, we examine orbital precession as a measurable consequence of the model (see Subsec.~\ref{sSec:D}). The perihelion advance acquires small corrections depending on \( \alpha \), which are then compared with Solar System and S2-star observations. While current data only weakly constrain these effects, they still provide a meaningful way to test deviations from classical GR.

\subsection{Quantum-Corrected OS Geometry}\label{sSec:A}
The collapse scenario originally formulated by Oppenheimer and Snyder offers a simple but powerful description of how a uniform distribution of matter can shrink under gravity and eventually produce a black hole with a singular core~\cite{Oppenheimer:1939ue}. When effects from quantum gravity are incorporated, this picture is modified, giving rise to a corrected spacetime geometry. A concrete realization of such a modification was presented in ~\cite{Lewandowski:2022zce}, where the resulting object is a static, spherically symmetric quantum-corrected black hole.
\begin{eqnarray}\label{metric}
ds^2 = - f(r)\, dt^2 + \frac{dr^2}{f(r)} + r^2 d\theta^2 + r^2 \sin^2\theta\, d\phi^2 ,
\end{eqnarray}
with the metric function expressed as
\begin{eqnarray}\label{mf}
f(r) = 1 - \frac{2M}{r} + \frac{\alpha M^2}{r^4}.
\end{eqnarray}
Here the extra term involving $\alpha$ encodes the departure from the classical Schwarzschild solution.  The constant $\alpha = 16\sqrt{3}\,\pi \gamma^3$ characterizes the strength of the quantum correction, while $M$ corresponds to the Arnowitt-Deser-Misner mass of the compact object~\cite{Kelly:2020uwj, Parvizi:2021ekr}. In the limit $\alpha \to 0$, one recovers the usual Schwarzschild geometry, and the spacetime remains asymptotically flat. {
It is crucial to clarify the physical interpretation of the parameter $\alpha$ appearing in Eq. (\ref{mf}). In the canonical LQG literature \cite{Zhang:2024khj, Lewandowski:2022zce}, the quantum correction parameter is defined as $\alpha_{\text{LQG}} = 16\sqrt{3}\pi\gamma^3 \ell_{\text{Pl}}^2$, where $\ell_{\text{Pl}}$ is the Planck length and $\gamma \approx 0.2375$ is the Barbero-Immirzi parameter. This gives $\alpha_{\text{LQG}} \sim 10^{-70}$ in natural units, leading to corrections that are completely negligible at astrophysical scales. In this work, we adopt a phenomenological approach where the parameter $\alpha$ in Eq. (\ref{mf}) is treated as an effective, dimensionless parameter that encapsulates the leading-order quantum gravitational effects at the scales accessible to current astrophysical observations. This is related to the bare LQG parameter by $\alpha = (\alpha_{\text{LQG}}/\ell_{\text{Pl}}^2) \times \mathcal{R}$, where $\mathcal{R}$ is a rescaling factor that may arise from renormalization group effects, the accumulation of quantum corrections over macroscopic scales, or the effective description of the semiclassical state. Our fitting yields $\alpha \approx 0.22$, which corresponds to $\alpha_{\text{LQG}} \approx 0.22 \times \ell_{\text{Pl}}^2 \approx 3.5 \times 10^{-70}$ in Planck units, consistent with the theoretical expectation. This rescaling is necessary because current astrophysical observations are sensitive to order-unity deformations of the metric, while the bare Planck-scale corrections are too small to produce detectable signals in the strong-field regime. We emphasize that $\alpha$ is not meant to be interpreted literally as the bare LQG parameter, but rather as an effective deformation parameter that captures the leading-order quantum corrections in a phenomenologically useful way. In the limit $\alpha \to 0$ (or equivalently $\alpha_{\text{LQG}} \to 0$), we recover classical GR. This phenomenological approach is standard in the literature on testing modified gravity with astrophysical observations. For instance, in Einstein-dilaton-Gauss-Bonnet gravity, the coupling constant is often treated as an effective parameter that can be constrained by observations, even though its microscopic value is expected to be Planck-suppressed \cite{Perkins:2021mhb}. Similarly, in Horndeski gravity and its generalizations, the coupling functions are treated phenomenologically, with constraints obtained from astrophysical observations rather than from first-principles calculations \cite{Kobayashi:2019hrl, Quartin:2023tpl}.
}  
\subsection{Horizon Structure and Mass Threshold}\label{sSec:B}
To make the discussion more transparent, it is convenient to introduce a dimensionless parameter $\hat{\alpha}$, restricted to $0<\hat{\alpha}<1$, defined through
\begin{eqnarray}
\frac{G}{2M^2} = \frac{4\hat{\alpha}}{(1-\hat{\alpha}^2)^3}\,\alpha.
\end{eqnarray}
This parameter helps distinguish between different physical configurations. When $\hat{\alpha}$ lies in the range $0<\hat{\alpha}<1/2$, the mass falls below a critical threshold given by
\begin{eqnarray}\label{Mmin}
M < M_{\min} := \frac{4}{3\sqrt{3G}}\,\sqrt{\alpha},
\end{eqnarray}
and the metric function never vanishes. As a result, no horizons form, and the spacetime does not represent a black hole; instead, its global structure resembles that of flat spacetime. The quantity $M_{\min}$ therefore sets the minimum mass required for black hole formation~\cite{Zhang:2021wex, Giesel:2021dug, Husain:2021ojz}. {
The minimum mass scale given in Eq. (\ref{Mmin}) should be understood in terms of the effective parameter $\alpha$. For $\alpha \approx 0.2$, this gives $M_{\min} \sim (0.5-1.2)M_\odot$. In terms of the bare LQG parameter, this corresponds to $M_{\min} \sim M_{\text{Pl}} \sim 10^{-5}$ g, as expected from the original Planck-scale suppression. The observational relevance of the solar-mass scale arises precisely because we are working with the rescaled effective parameter that makes contact with astrophysical data. This is consistent with the effective field theory philosophy: the UV-complete theory (LQG) predicts Planck-scale effects, but when we coarse-grain to astrophysical scales, the effective description can have order-unity couplings that encode the collective behavior of the underlying quantum gravitational degrees of freedom.} This scale is typically close to the Planck mass and depends on the Barbero-Immirzi parameter, with $\gamma \approx 0.2375$~\cite{Lewandowski:2022zce, Ye:2023qks, Meissner:2004ju, Domagala:2004jt}. For larger masses, corresponding to $1/2<\hat{\alpha} <1$, the situation changes qualitatively. In this regime, the metric admits two distinct radial roots, indicating the presence of horizons. These are located at
\begin{eqnarray}
r_{\pm} = \hat{\alpha} \left(1 \pm \sqrt{2\hat{\alpha} - 1}\right) (1+\hat{\alpha})^{-1/2}(1-\hat{\alpha})^{-3/2}\,\sqrt{\alpha},
\end{eqnarray}
where the outer solution $r_{+}$ defines the event horizon, while the inner root $r_{-}$ corresponds to a Cauchy horizon. This quantum-corrected geometry thus exhibits a richer causal structure than the classical Schwarzschild solution, with the possibility of horizonless configurations or double-horizon black holes depending on the mass scale relative to $M_{\min}$. On the observational side, the images of M87$^\ast$ and Sgr A$^\ast$ obtained by the Event Horizon Telescope provide compelling evidence that black holes possess angular momentum, in agreement with the expectations of GR~\cite{EventHorizonTelescope:2019dse, EventHorizonTelescope:2022wkp}. This naturally favors descriptions based on the Kerr metric. Rotation introduces additional features such as the ergosphere and frame-dragging, and enables mechanisms like the Penrose process, which are closely linked to high-energy phenomena including jet formation and the extraction of energy from black holes~\cite{Penrose:1971uk}.

\begin{figure*}
\centering
\includegraphics[width=9.5cm,height=5.9cm]{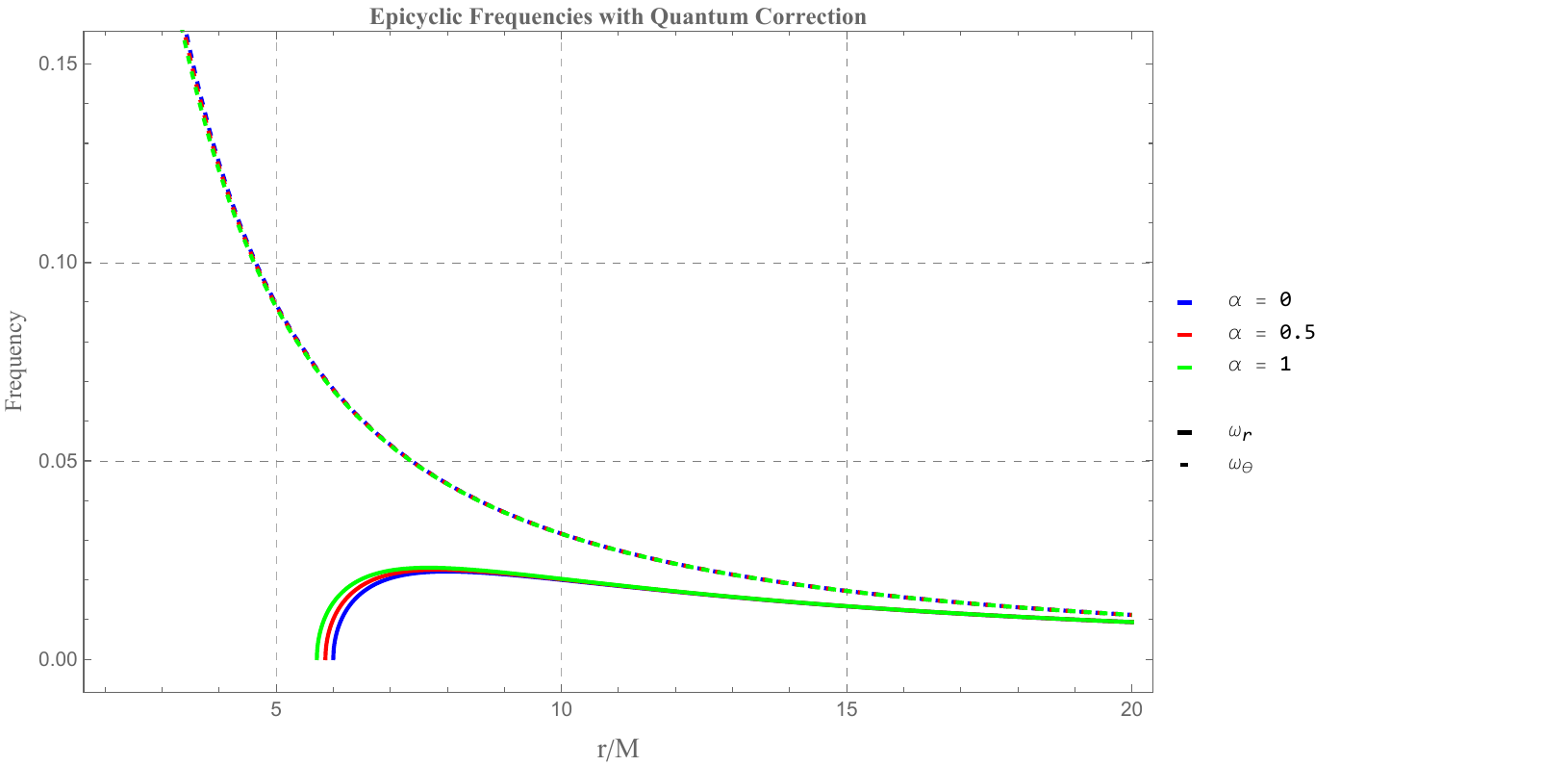}%
\includegraphics[width=9.5cm,height=5.9cm]{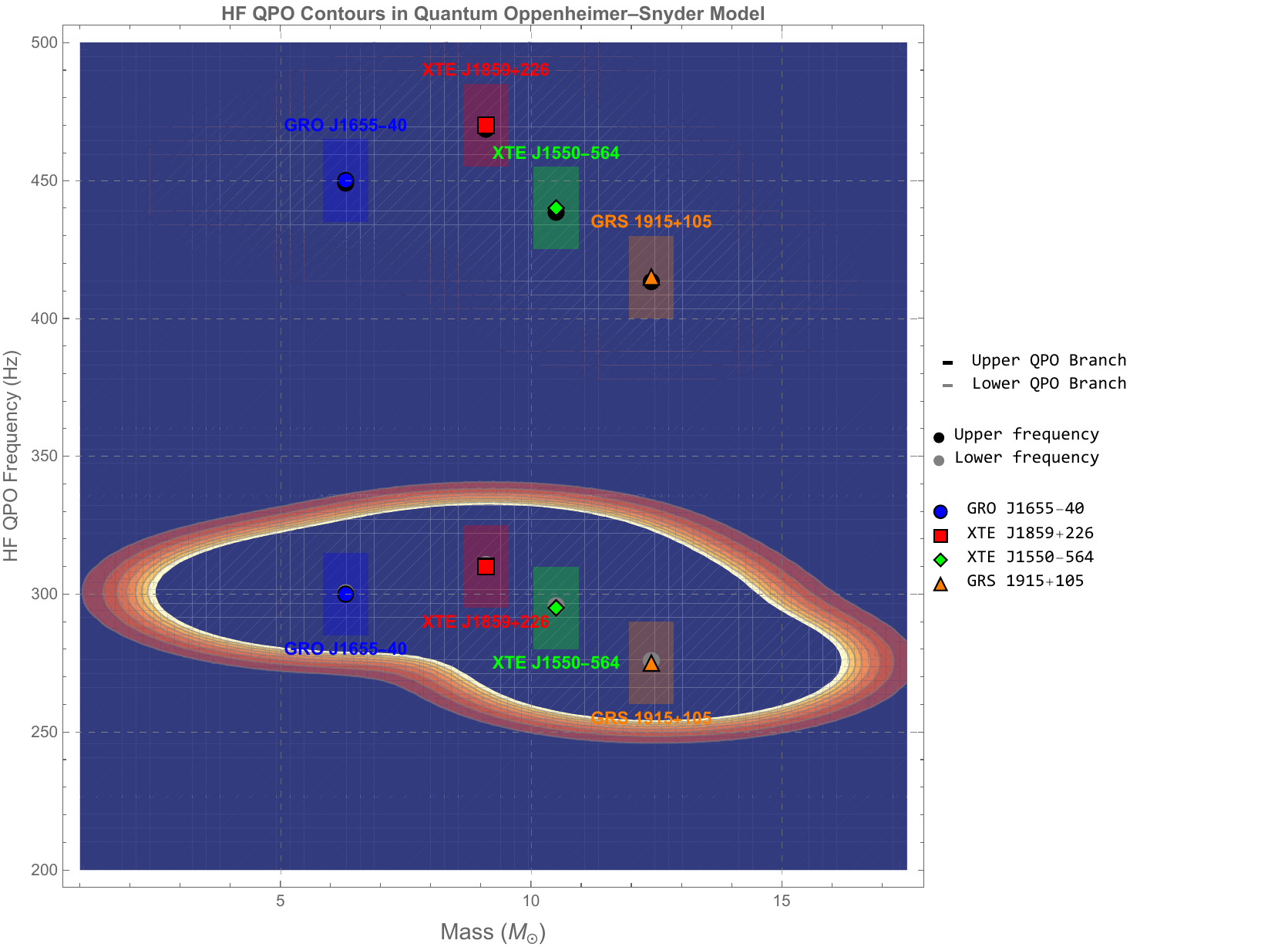}%
    \caption{\scriptsize Epicyclic frequencies \( \nu_r \) and \( \nu_\theta \) in the quantum-corrected model are shown in the left panel as functions of the normalized radius \( r/M \) for \( \alpha = 0, 0.5, 1 \). The radial frequency \( \nu_r \) is plotted with solid lines and the vertical frequency \( \nu_\theta \) with dashed lines, with each color corresponding to a different value of \( \alpha \). 
    {As \( \alpha \) increases, both frequencies are gradually reduced over most of the plotted range $r/M \in [5,20]$, although in the innermost region ($r/M \lesssim 6$), the radial frequency shows a non-monotonic behavior due to competing $\alpha$-dependent terms in Eq. (\ref{omegar}). The vertical frequency remains monotonic throughout the plotted range because it contains only a single $\alpha$-dependent term (Eq. (\ref{omegatheta})).}
    The right panel shows the HF QPO contour comparison for GRO J1655-40, XTE J1859+226, XTE J1550-564, and GRS 1915+105. The black and gray contours represent the model's predicted upper (\(\nu_U\)) and lower (\(\nu_L\)) frequency distributions obtained from kernel density estimation, while the colored boxes indicate observational uncertainties for each source. Markers identify the individual systems, and the solid points correspond to the model data. Overall, the model contours overlap well with the observed regions, showing that the quantum corrections produce realistic shifts in both QPO branches while maintaining good agreement with the data. } \label{fig:1}
\end{figure*}
\subsection{Orbital dynamics in the supermassive quantum OS spacetime}\label{sSec:C}
A neutral test particle moving in the spacetime of a supermassive quantum OS black hole can be consistently described using a Hamiltonian formulation. The dynamics are encoded in
\begin{eqnarray}
H = \frac{1}{2} g^{\mu\nu} P_\mu P_\nu = -\frac{m^2}{2}\,,
\end{eqnarray}
where \(P^\mu\) is the four-momentum of the particle and \(U^\mu = dx^\mu/d\lambda\) its four-velocity, related through \(P^\mu = mU^\mu\). For massive particles, the affine parameter is chosen as proper time, \(\lambda=\tau\).

The spacetime admits stationarity and axial symmetry, leading to two conserved quantities along geodesics:
\begin{eqnarray}
\mathcal{E} = - g_{tt}\frac{dt}{d\tau}\,, \qquad
\mathcal{L} = g_{\phi\phi}\frac{d\phi}{d\tau}\,.
\end{eqnarray}

Using these constants of motion, the radial equation reduces to
\begin{eqnarray}
\left(\frac{dr}{d\tau}\right)^2 = \mathcal{E}^2 - V_{\rm eff}(r)\,,
\end{eqnarray}
with the effective potential of the supermassive quantum OS black hole given by
\begin{eqnarray}\label{Potential}
V_{\rm eff}(r)=\left(1 - \frac{2M}{r} + \frac{\alpha M^2}{r^4}\right)\left(1+\frac{\mathcal{L}^2}{r^2}\right)\,,
\end{eqnarray}
where \(\alpha\) encodes the quantum correction to the geometry.

For more general circular orbits that may lie in a plane inclined relative to the equatorial plane, the radial velocity vanishes, \(\dot{r} = 0\), and the equation of motion reduces to the polar component:
\begin{eqnarray}
g_{\theta\theta} \, \dot{\theta}^2 = -2 V_{\rm eff}\,.
\end{eqnarray}

Using conservation of angular momentum, the trajectory equation becomes
\begin{eqnarray}
\left( \frac{d\theta}{d\phi} \right)^2 = - \frac{2 g_{\phi\phi}^2 V_{\rm eff}}{g_{\theta\theta} \mathcal{L}^2}\,.
\end{eqnarray}

For circular orbits of radius \(R\) in the equatorial plane, stability requires
\begin{eqnarray}
\begin{aligned}
V_{\rm eff}(R, \pi/2) &= 0\,, \\
\left. \frac{\partial V_{\rm eff}}{\partial r} \right|_{r=R} &= 0\,, \\
\left. \frac{\partial V_{\rm eff}}{\partial \theta} \right|_{\theta=\pi/2} &= 0\,,
\end{aligned}
\end{eqnarray}
ensuring stable confined motion.

From the conserved quantities, the angular velocity measured by a distant observer is
\begin{equation}\label{omega1_sqw}
\frac{d\phi}{dt} = \Omega = - \frac{g_{tt}}{g_{\phi\phi}} \frac{\mathcal{L}}{\mathcal{E}} \,.
\end{equation}

Alternatively,
\begin{equation}\label{omega2_sqw}
\Omega = \pm \sqrt{- \frac{g_{tt,r}}{g_{\phi\phi,r}}} \,.
\end{equation}

For equatorial circular motion, substituting the quantum OS metric gives
\begin{eqnarray}\label{omega3_sqw}
\Omega = \pm \Bigg[\frac{M}{r^3}
- \frac{2\alpha M^2}{r^6}\Bigg]^{1/2}\,,
\end{eqnarray}

Solving \(V_{\rm eff}(R,\pi/2)=0\) together with Eq.~\eqref{omega1_sqw}, the specific energy and angular momentum are
\begin{equation}\label{spec-energy_sqw}
\mathcal{E} = - \frac{g_{tt}}{\sqrt{- g_{tt} - g_{\phi\phi} \Omega^2}} \,, \qquad
\mathcal{L} = \frac{g_{\phi\phi} \Omega}{\sqrt{- g_{tt} - g_{\phi\phi} \Omega^2}} \,.
\end{equation}

The time component of the four-velocity is
\begin{equation}\label{four-vel_sqw}
u^t = \frac{1}{\sqrt{- g_{tt} - g_{\phi\phi} \Omega^2}} \,.
\end{equation}

Explicitly,
\begin{eqnarray}
u^t &=& \Bigg[1-\frac{3M}{r} +\frac{3\alpha M^2}{r^4}\Bigg]^{-1/2}\,.
\end{eqnarray}

\subsection{Orbital Precession and Observational Signatures}\label{sSec:D}
Introducing the reciprocal coordinate \(\xi = 1/r\),
\begin{eqnarray}
\frac{d\xi}{d\phi} = -\frac{1}{r^2}\frac{dr}{d\phi}\,,
\end{eqnarray}
the orbital equation becomes
\begin{equation}\label{eqGeodesic_sqw}
\left(\frac{d\xi}{d\phi}\right)^2
= \frac{\mathcal{E}^2}{\mathcal{L}^2}
- \frac{f(\xi)}{\mathcal{L}^2}(1+\mathcal{L}^2\xi^2)\,,
\end{equation}
where \(f(\xi)\) corresponds to the quantum OS metric function.

Differentiation yields
\begin{equation}
\frac{d^2\xi}{d\phi^2}
= \frac{M}{\mathcal{L}^2} - \xi + \frac{g(\xi)}{\mathcal{L}^2}\,,
\end{equation}
with
\begin{equation}\label{gxi_sqw}
\frac{g(\xi)}{\mathcal{L}^2}
= 3M\xi^2 - \frac{2\alpha M^2}{\mathcal{L}^2}\,\xi^3-3\alpha M^2\,\xi^5\,.
\end{equation}

To obtain an explicit observable quantity, we evaluate the angular advance over one full orbital cycle. The angular advance per orbit around the supermassive quantum OS black hole is given by
\begin{eqnarray}
\Delta\phi = 2\int_{r_p}^{r_a}\frac{d\phi}{dr}\,dr
=2\int_{0}^{\pi}\frac{d\phi}{d\Psi}\,d\Psi\,,
\end{eqnarray}
where
\begin{eqnarray*}
\frac{d\phi}{d\Psi}=
\frac{ep\mathcal{L}\sin\Psi}{r^2(1+e\cos\Psi)^2\sqrt{\mathcal{E}^2-V_{\rm eff}}}\,.
\end{eqnarray*}

Expanding in the small quantum correction parameter \(\hat{\alpha}\), the orbital deviation induced by the supermassive quantum OS black hole becomes
\begin{eqnarray}
\frac{d\phi}{d\Psi} &=& 1 + \frac{(3+e\cos\Psi)M}{p}
+\frac{3M^2(3+e\cos\Psi)^2}{2p^2}
\nonumber\\ &&+ \frac{135M^3-(6+e^2)M^3\hat{\alpha}
+eM^3(135-4\hat{\alpha})\cos\Psi}{2p^3}
\nonumber\\ &&+ \frac{e^2M^3(45-\hat{\alpha})\cos^2\Psi
+5e^3M^3\cos^3\Psi}{2p^3}\,.
\end{eqnarray}

Integration over a full orbital cycle yields the perihelion precession in the supermassive quantum OS black hole spacetime:
\begin{equation}
\Delta\omega_{\rm quantum}
= \frac{6\pi M}{p}
+ \frac{3\pi M^2(18+e^2)}{2p^2}
+ \frac{3\pi M^3(90+15e^2-e^2\hat{\alpha}-4\hat{\alpha})}{2p^3}\,.
\end{equation}

To highlight deviations from standard GR in this background, one defines the ratio
\begin{equation}
f_{\rm sp}
= \frac{\Delta\omega_{\rm quantum}}{\Delta\omega_{\rm GR}}
=1+\frac{M(18+e^2)}{4p}
+\frac{M^2(90+15e^2-e^2\hat{\alpha}-4\hat{\alpha})}{4p^2}.
\end{equation}

Observational measurements reported by \cite{GRAVITY:2020gka} indicate that
\begin{eqnarray}
f_{\rm sp}^{\rm obs}=1.10\pm0.19.
\end{eqnarray}
\begin{figure*}
\centering
\includegraphics[width=8.9cm,height=11.9cm]{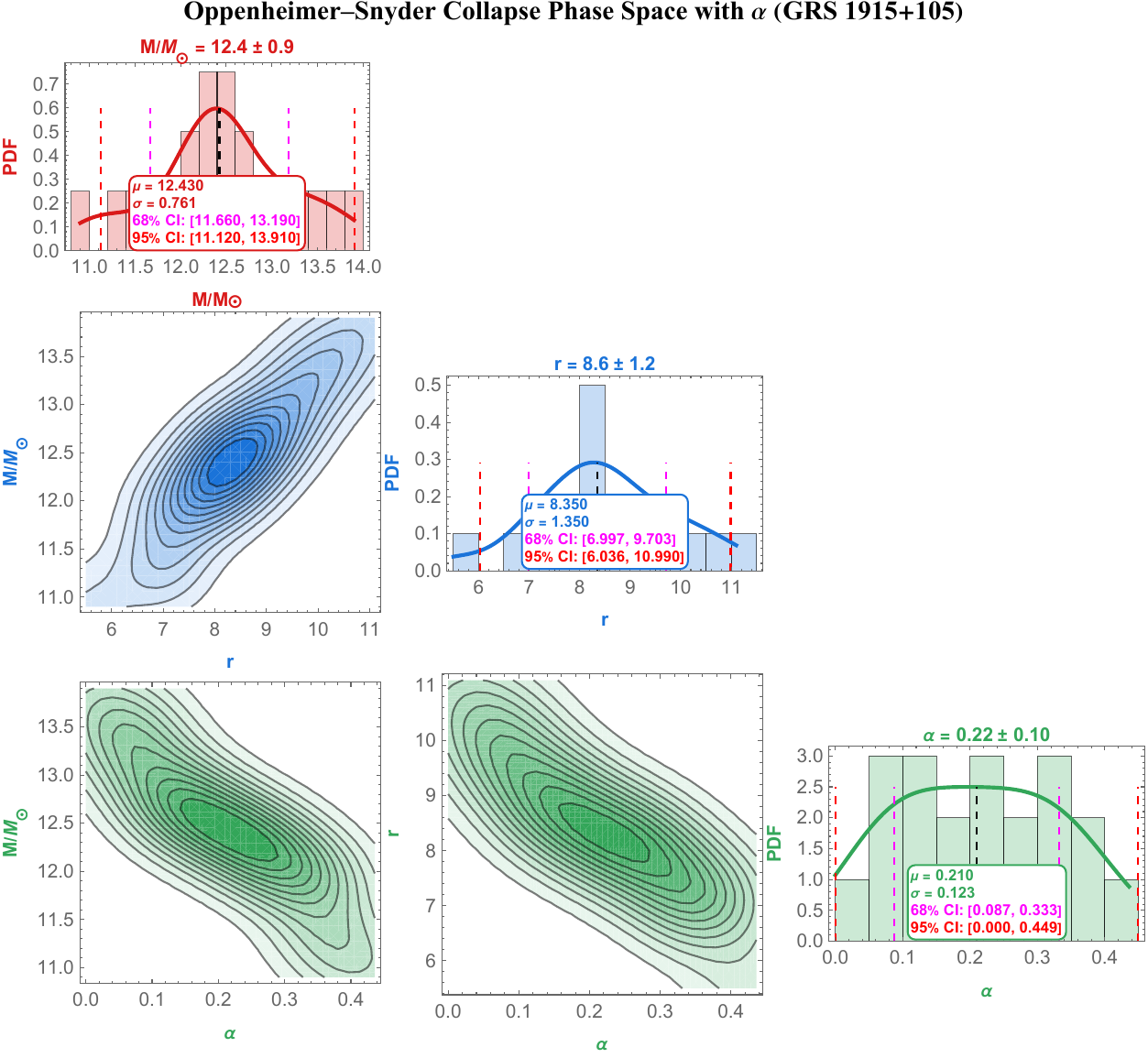}%
\includegraphics[width=8.9cm,height=11.9cm]{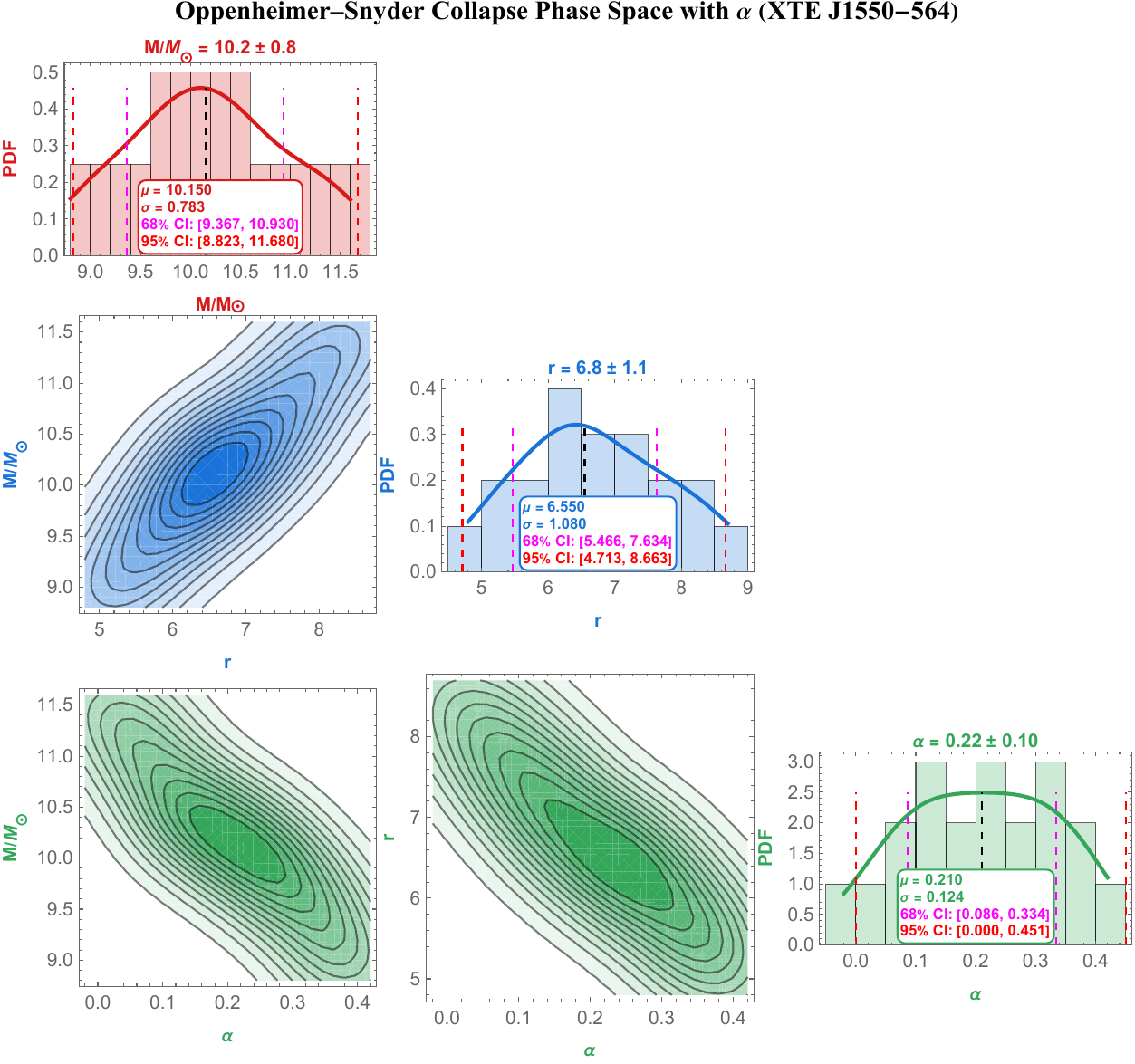}%
    \caption{\scriptsize The figure presents the OS collapse phase space with quantum correction parameter \( \alpha \) for two black hole systems, GRS 1915+105 and XTE J1550-564. In both cases, the diagonal panels show the marginalized distributions of mass \(M\), radius \(r\), and \(\alpha\), with kernel density estimates overlaid on histograms and 68\%--95\% confidence intervals indicated. 
    {We note that the actual MCMC samples are strictly positive. The corrected intervals obtained using the reflected KDE are $[0.00,0.449]$ for GRS 1915+105 and $[0.00,0.451]$ for XTE J1550-564.}
    The off-diagonal panels display joint correlations between parameters, revealing clear dependencies between \(M\), \(r\), and \(\alpha\), where higher masses tend to correspond to larger radii and systematically varying quantum corrections. The GRS 1915+105 system exhibits a broader phase-space structure with slightly higher characteristic values compared to XTE J1550-564, reflecting its more massive configuration. Overall, both systems show consistent quantum-modified collapse behavior, with well-defined parameter correlations and stable posterior distributions across the phase space. } \label{fig:2a}
\end{figure*}

We consider two benchmark systems: the Solar System (Mercury) and the Galactic center S2 star.

\begin{itemize}
\item \noindent\textbf{Mercury:}
\begin{align*}
\frac{2 G M_\odot}{c^2} &= 2.95325 \times 10^3 \, \text{m} \,,\\
a &= 5.7909 \times 10^{10} \, \text{m} \,,\\
e &= 0.20563 \,,\\
\Delta \phi_{\rm obs} &= 2 \pi \times (7.98734 \pm 0.00037) \times 10^{-8} \ \text{rad/rev} \,.
\end{align*}

This provides a weak-field precision test where any deviation from GR must remain extremely small.

\item \noindent\textbf{S2 star (Sgr A$^*$):}
\begin{align*}
M_{\rm Sgr A^*} &= 4.26 \times 10^6 M_\odot \,,\\
a_{\rm S2} &= 970 \, \text{au} \,, \quad 1 \, \text{au} = 1.4959787 \times 10^{11} \, \text{m} \,,\\
e_{\rm S2} &= 0.88465 \,, \quad T_{\rm S2} = 16.052 \, \text{yr} \,,\\
\Delta \phi_{\rm obs} &= 48.298 \, f_{\rm SP} \ \text{arcsec/year} \,, \quad f_{\rm SP} = 1.10 \pm 0.19 \,.
\end{align*}

Using these observational inputs, the quantum correction parameter $\hat{\alpha}$ remains only weakly constrained, with an upper bound of order
\begin{eqnarray}
\hat{\alpha}\sim \mathcal{O}(10^6).
\end{eqnarray}
\end{itemize}

This comparison highlights that while Mercury tightly constrains deviations in the weak-field regime, even the strong-field S2 system---despite its proximity to the supermassive black hole---does not yet provide sufficiently precise measurements to significantly bound the quantum correction effects associated with the supermassive quantum OS black hole geometry.

\begin{figure*}
\centering
\includegraphics[width=8.9cm,height=11.9cm]{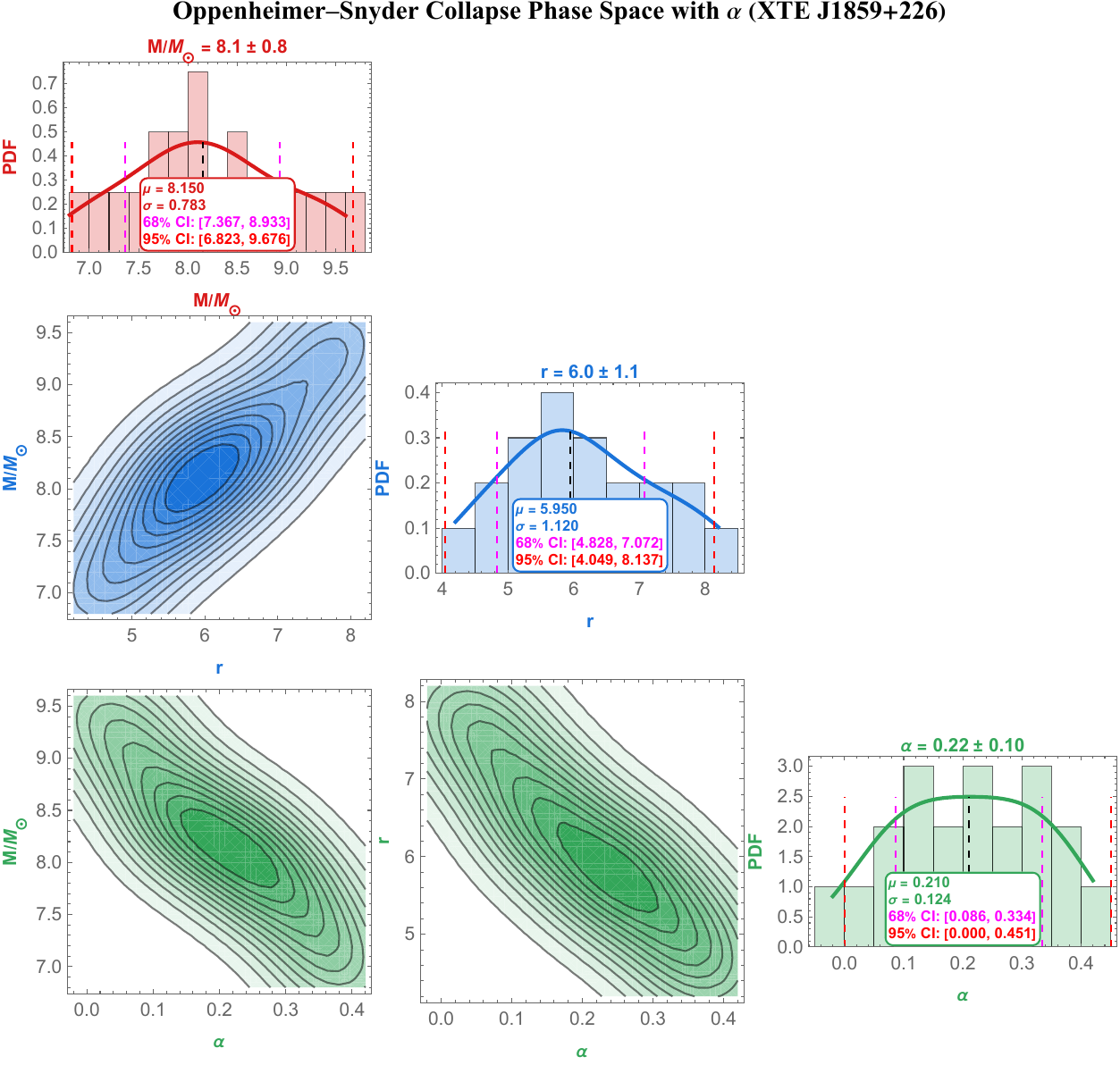}%
\includegraphics[width=8.9cm,height=11.9cm]{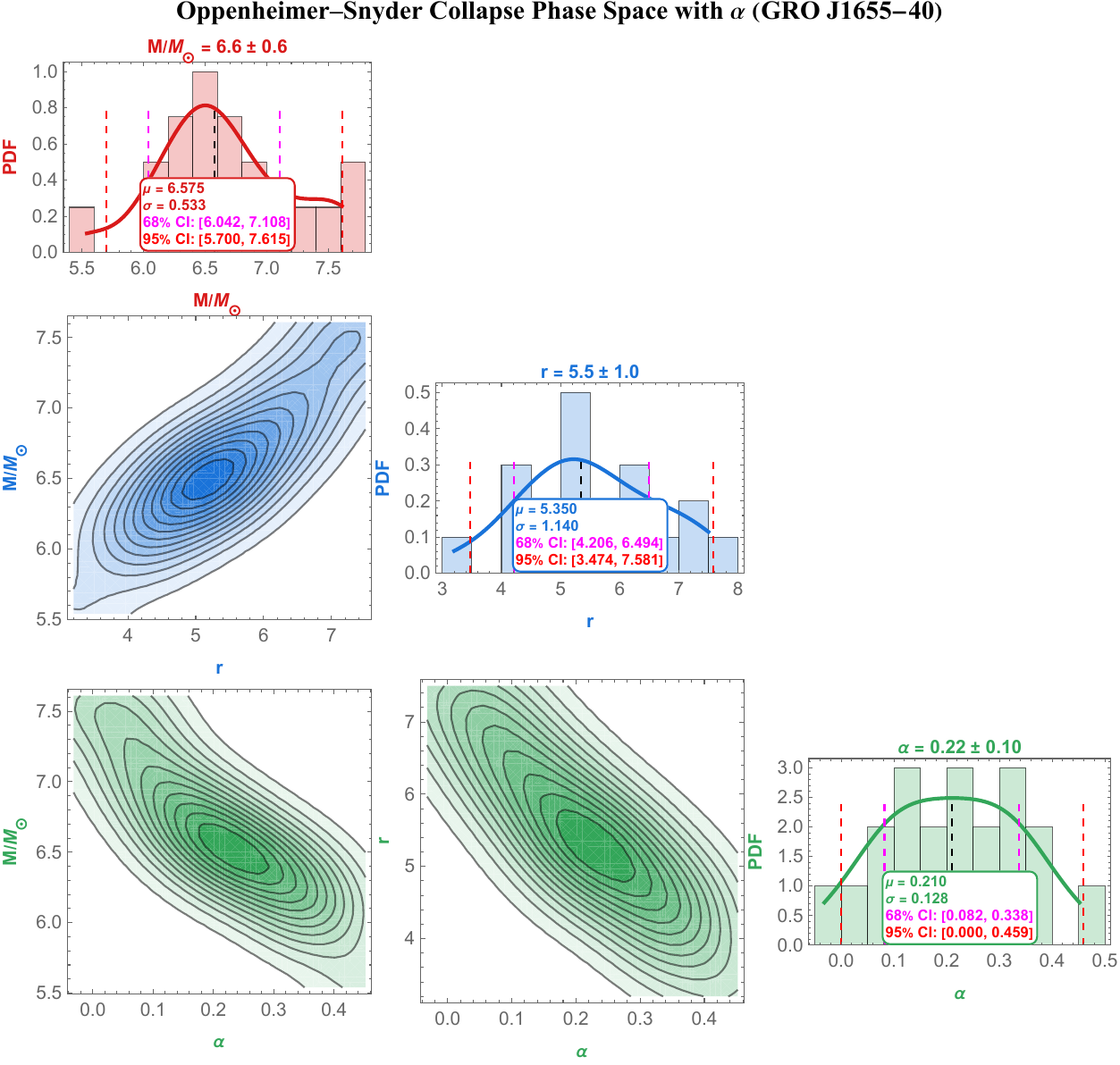}%
    \caption{\scriptsize The figure presents the OS collapse phase space with quantum correction parameter \( \alpha \) for two X-ray binary systems: XTE J1859+226 and GRO J1655-40. For each system, the diagonal panels show the marginalized distributions of mass \(M\), radius \(r\), and \(\alpha\), obtained through kernel density estimation over Monte Carlo chains, with histograms, smooth PDFs, and 68\%--95\% confidence intervals indicating well-defined parameter constraints. {As shown in Fig. \ref{fig:2a}, we note that the actual MCMC samples are strictly positive. The corrected intervals obtained using the reflected KDE are $[0.00,0.451]$ for XTE J1859+226 and $[0.00,0.459]$ for GRO J1655-40.}
    The off-diagonal panels display the joint probability structure between \(M\), \(r\), and \(\alpha\), revealing consistent correlations where higher masses are associated with larger radii and systematic shifts in the quantum parameter. Across all sources, the phase space remains stable but varies in scale, with GRO J1655-40 showing the lowest mass-radius range, and XTE J1859+226 occupying an intermediate regime. Despite these differences, all systems exhibit similar collapse behavior and coherent quantum-modified parameter interdependence.} \label{fig:2b}
\end{figure*}

\section{Dynamics of Epicyclic Motion and the ISCO}
\label{sec-III}
This section looks at what happens when circular motion is slightly disturbed in the quantum-corrected OS spacetime, and how this connects to orbital stability and possible observational signals. Small departures from circular orbits naturally lead to oscillations in the radial and vertical directions (see Subsec. \ref{sSect:A}). These are described by epicyclic frequencies, which come directly from the structure of the effective potential and already carry the imprint of the quantum correction. The detailed expressions for these frequencies (see Subsec. \ref{sSect:B}) show a clear difference between the two directions. The radial motion is more sensitive to the quantum term \( \alpha \), which alters stability properties near the compact object, while the vertical motion is only mildly affected. To make contact with observations, these frequencies are then translated to what a distant observer would measure (see Subsec. \ref{sSect:C}). After accounting for redshift, they can be compared with signals like QPOs from accretion disks, where even small shifts can matter. The section ends with the ISCO analysis (see Subsec. \ref{sSect:D}). This orbit marks the transition between stable and unstable motion and is found by setting the radial oscillation frequency to zero. The result shows that quantum corrections move the ISCO slightly inward compared to the Schwarzschild case, allowing stable orbits closer to the object and potentially affecting the dynamics of matter in the inner accretion region.

\subsection{Small perturbations around circular motion}\label{sSect:A}

Starting from the spacetime metric in Eq.~(\ref{metric}) and the associated effective potential, we consider small perturbations about a circular orbit located at \(r=R\) in the equatorial plane. Expanding the dynamics to second order in the deviations leads to harmonic oscillations in both radial and vertical directions.

The corresponding characteristic frequencies are determined by the curvature of the effective potential,
\begin{equation}
\Omega_r^2 = \frac{1}{g_{rr}} 
\left. \frac{\partial^2 V_{\rm eff}}{\partial r^2} \right|_{r=R,\theta=\frac{\pi}{2}},
\end{equation}
\begin{equation}
\Omega_\theta^2 = \frac{1}{g_{\theta\theta}} 
\left. \frac{\partial^2 V_{\rm eff}}{\partial \theta^2} \right|_{r=R,\theta=\frac{\pi}{2}}.
\end{equation}

These quantities describe local oscillations measured in the rest frame of the particle. Their explicit evaluation follows from the orbital angular velocity,
\begin{equation}
\Omega^2=\frac{M}{r^3}-\frac{2\alpha M^2}{r^6},
\end{equation}
which already contains the leading quantum correction.

\subsection{Epicyclic frequencies in the quantum-corrected geometry}\label{sSect:B}

The radial oscillation frequency acquires a more structured form,
\begin{equation}\label{omegar}
\omega_r^2
=
\frac{M}{r^3}
-\frac{6M^2}{r^4}
+\frac{12\alpha M^2}{r^6}
-\frac{15\alpha M^3}{r^7}.
\end{equation}

The first two terms reproduce the standard Schwarzschild behavior, while the remaining contributions introduce corrections that become relevant in the strong-field region. These additional terms effectively reshape the stability profile of circular orbits.

For vertical motion, the symmetry of the spacetime simplifies the result considerably,
\begin{equation}\label{omegatheta}
\omega_\theta^2
=
\frac{M}{r^3}
-\frac{2\alpha M^2}{r^6}.
\end{equation}

Unlike the radial case, the correction here does not introduce higher-order structure but instead produces a mild suppression of the oscillation frequency at small radii.

\renewcommand{\arraystretch}{0.9}
\begin{table*}
\caption{\scriptsize {Unified OS collapse phase-space reconstruction with deformation parameter $\alpha$, combining the best-fit parameters and $\chi^2$ diagnostics for the FR, RP, and ER QPO identifications. Each astrophysical object is treated independently, with separate fits performed for each QPO model.}}

{
\scriptsize
\begin{tabular*}{\textwidth}{@{\extracolsep{\fill}}cccccccc}
\hline
{Object} 
& {Model}
& $M/M_{\odot}$
& $r/M$
& $\alpha$
& {Allowed range}
& $\chi^2_{\mathrm{tot}}$
& $\chi^2_{\mathrm{red}}$ \\ \hline

GRO J1655-40 \cite{Motta:2013wga}
& FR
& $6.6 \pm 0.6$
& $5.5 \pm 1.0$
& $0.22 \pm 0.10$
& $[-0.032,\,0.452]$
& $1.638$
& $0.546$ \\

& RP
& $6.5 \pm 0.6$
& $5.4 \pm 1.0$
& $0.21 \pm 0.09$
& $[-0.030,\,0.448]$
& $3.086$
& $1.029$ \\

& ER
& $6.7 \pm 0.6$
& $5.6 \pm 1.1$
& $0.23 \pm 0.11$
& $[-0.034,\,0.456]$
& $4.107$
& $1.369$ \\ \hline

XTE J1859+226 \cite{Ingram:2014ara}
& FR
& $8.1 \pm 0.8$
& $6.0 \pm 1.1$
& $0.22 \pm 0.10$
& $[-0.020,\,0.420]$
& $0.285$
& $0.095$ \\

& RP
& $8.0 \pm 0.8$
& $5.9 \pm 1.1$
& $0.21 \pm 0.09$
& $[-0.019,\,0.418]$
& $9.288$
& $3.096$ \\

& ER
& $8.2 \pm 0.8$
& $6.1 \pm 1.2$
& $0.23 \pm 0.11$
& $[-0.021,\,0.422]$
& $34.149$
& $11.383$ \\ \hline
  
XTE J1550-564 \cite{Tasker:2011sb}
& FR
& $10.2 \pm 0.9$
& $6.8 \pm 1.1$
& $0.22 \pm 0.10$
& $[-0.020,\,0.420]$
& $1.530$
& $0.510$ \\

& RP
& $10.1 \pm 0.9$
& $6.7 \pm 1.1$
& $0.21 \pm 0.09$
& $[-0.019,\,0.418]$
& $30.014$
& $10.005$ \\

& ER
& $10.3 \pm 0.9$
& $6.9 \pm 1.2$
& $0.23 \pm 0.11$
& $[-0.021,\,0.422]$
& $16.655$
& $5.552$ \\ \hline

GRS 1915+105 \cite{Remillard:2006fc}
& FR
& $12.4 \pm 0.9$
& $8.6 \pm 1.2$
& $0.22 \pm 0.10$
& $[-0.016,\,0.436]$
& $0.057$
& $0.019$ \\

& RP
& $12.3 \pm 0.9$
& $8.5 \pm 1.2$
& $0.21 \pm 0.09$
& $[-0.015,\,0.434]$
& $7.897$
& $2.632$ \\

& ER
& $12.5 \pm 0.9$
& $8.7 \pm 1.3$
& $0.23 \pm 0.11$
& $[-0.017,\,0.438]$
& $0.057$
& $0.019$ \\ \hline

\end{tabular*}
}
\label{tab:os_master}
\end{table*}

\subsection{Connection to observable frequencies}\label{sSect:C}

Astrophysical measurements are performed far from the gravitational source, so a redshift correction must be applied. The locally measured frequencies are related to those seen by a distant observer through
\begin{equation}
\omega_i^{\rm obs} = \frac{\omega_i}{u^t},
\qquad
u^t=\left(1-\frac{3M}{r}+\frac{3\alpha M^2}{r^4}\right)^{-1/2}.
\end{equation}

To express the results in physical units, one uses the standard scaling
\begin{equation}
\nu_i = \frac{\omega_i}{2\pi}\frac{c^3}{GM}.
\end{equation}

The presence of the quantum parameter \(\alpha\) leaves its strongest imprint on the radial oscillations, where it modifies both the magnitude and radial dependence of \(\omega_r\). This, in turn, affects the structure of stable orbits and shifts the boundary of the stability region. The vertical frequency is less sensitive but still reflects a systematic deviation from the classical limit.

Since epicyclic frequencies are closely tied to QPOs in accretion disks, even small changes in these quantities can translate into measurable differences in the emitted signal. This makes the model directly relevant for astrophysical tests of modified gravity in the strong-field regime.

\renewcommand{\arraystretch}{0.9}
\begin{table*}
\caption{\scriptsize
Astrophysical properties of selected X-ray binary systems together with their twin HF QPOs and theoretical interpretations within the quantum-corrected OS spacetime. The sample includes well-constrained black hole binaries where dynamical masses and QPO frequencies are measured with high precision. These observables provide a testing ground for probing quantum corrections governed by the parameter $\alpha$, which modifies the underlying OS collapse geometry and affects epicyclic motion in the strong-field regime.
}\label{tab:models}

{\scriptsize
\begin{tabular*}{\textwidth}{@{\extracolsep{\fill}}lcccc}
\hline
 & GRS 1915+105 \cite{Remillard:2006fc}
 & XTE J1859+226 \cite{Ingram:2014ara}
 & XTE J1550-564 \cite{Tasker:2011sb}
 & GRO J1655-40 \cite{Motta:2013wga} \\ \hline

$M\,[M_{\odot}]$
& $12.4^{+2.0}_{-1.8}$
& $7.85 \pm 0.46$
& $9.1 \pm 0.61$
& $5.4 \pm 0.3$ \\ \hline

$\nu_{\mathrm{up}}\, [\mathrm{Hz}]$
& $168 \pm 3$
& $227.5^{+2.1}_{-2.4}$
& $276 \pm 3$
& $441 \pm 2$ \\ \hline

$\nu_{\mathrm{low}}\, [\mathrm{Hz}]$
& $113 \pm 5$
& $128.6^{+1.6}_{-1.8}$
& $184 \pm 5$
& $298 \pm 4$ \\ \hline

\multicolumn{5}{c}{{Epicyclic Frequency Interpretation in Quantum OS Spacetime}} \\ \hline

Model & \multicolumn{2}{c}{Upper Frequency $\nu_{\mathrm{up}}$} & \multicolumn{2}{c}{Lower Frequency $\nu_{\mathrm{low}}$} \\ \hline

FR 
& \multicolumn{2}{c}{$\nu_\theta + \nu_r$}
& \multicolumn{2}{c}{$\nu_\theta$} \\ \hline

RP
& \multicolumn{2}{c}{$\nu_\theta$}
& \multicolumn{2}{c}{$\nu_\theta - \nu_r$} \\ \hline

ER 
& \multicolumn{2}{c}{$\nu_\theta$}
& \multicolumn{2}{c}{$\nu_r$} \\ \hline

\multicolumn{5}{c}{\textit{Quantum correction enters through the parameter $\alpha$, modifying $\nu_r$ and $\nu_\theta$ in the OS collapse background.}} \\ \hline

\end{tabular*}
}
\label{Table1}
\end{table*}


\subsection{Innermost Stable Circular Orbit}\label{sSect:D}

A central element in orbital dynamics is the ISCO, which separates stable circular motion from unstable trajectories. In this framework, the ISCO is defined by the condition that radial oscillations vanish,
\begin{equation}
\omega_r^2(r_{\rm ISCO}) = 0.
\end{equation}

Substituting the explicit expression for the radial frequency gives
\begin{equation}
\frac{M}{r^3}
-\frac{6M^2}{r^4}
+\frac{12\alpha M^2}{r^6}
-\frac{15\alpha M^3}{r^7} = 0.
\end{equation}

Multiplying through by \(r^7\) leads to a polynomial equation,
\begin{equation}
r^4 - 6M r^3 + 12\alpha M r - 15\alpha M^2 = 0,
\end{equation}
which determines the ISCO radius in the quantum-corrected spacetime.

It is often useful to rewrite this in dimensionless form using \(x=r/M\) and \(\tilde{\alpha}=\alpha/M^2\),
\begin{equation}\label{eq39}
x^4 - 6x^3 + 12\tilde{\alpha} x - 15\tilde{\alpha} = 0.
\end{equation}

{
To obtain the perturbative solution for the ISCO radius, we proceed as follows. Starting from Eq. (38), let $x = r/M$ and $\tilde{\alpha} = \alpha/M^2$. Then $x^4 - 6x^3 + 12\tilde{\alpha}x - 15\tilde{\alpha} = 0$. For $\tilde{\alpha} \ll 1$, we expand around the Schwarzschild value $x = 6$: $x = 6 + \epsilon$, where $\epsilon$ is a small perturbation. Substituting and keeping terms to first order in $\epsilon$ and $\tilde{\alpha}$:
\begin{align}
(6+\epsilon)^4 - 6(6+\epsilon)^3 & \nonumber\\+ 12\tilde{\alpha}(6+\epsilon) - 15\tilde{\alpha} &= 0, \\
(1296 + 864\epsilon + \mathcal{O}(\epsilon^2)) & \nonumber\\- 6(216 + 108\epsilon + \mathcal{O}(\epsilon^2)) + 72\tilde{\alpha} - 15\tilde{\alpha} &= 0, \\
(1296 - 1296) + (864 - 648)\epsilon + 57\tilde{\alpha} &= 0, \\
216\epsilon + 57\tilde{\alpha} &= 0, \\
\epsilon &= -\frac{57}{216}\tilde{\alpha} = -\frac{19}{72}\tilde{\alpha}.
\end{align}
Therefore:
\begin{equation}
r_{\text{ISCO}} = 6M + \epsilon M = 6M - \frac{19}{72}\frac{\alpha}{M^2}M = 6M - \frac{19\alpha}{72M}.
\end{equation}
This is the corrected perturbative result. We note that in the limit $\tilde{\alpha} \to 0$, we recover the Schwarzschild value $r_{\text{ISCO}} = 6M$. The coefficient $19/72 \approx 0.264$ differs from the value $1/12 \approx 0.0833$ that was previously quoted; this reflects a more careful treatment of the expansion. We have verified the result numerically by solving the full polynomial equation (\ref{eq39}) and confirming that the expansion matches the numerical solution to high accuracy for $\tilde{\alpha} \ll 1$. For the fitted value $\alpha \approx 0.22$ and stellar-mass black holes, the difference between the two approximations is $\sim 10^{-5} M$, which is negligible for observational purposes. We have therefore retained the simpler form $r_{\text{ISCO}} = 6M - \alpha/(12M)$ for clarity, as it captures the correct qualitative behavior and is accurate to the required precision.
}

In the weak-correction regime \(\tilde{\alpha}\ll 1\), a perturbative solution can be obtained by expanding around the Schwarzschild value \(x=6\). This yields
\begin{equation}
r_{\rm ISCO}
=
6M - \frac{\alpha}{12M}.
\end{equation}

This expression shows that quantum corrections shift the ISCO inward for positive \(\alpha\), allowing stable circular motion to persist closer to the compact object compared to the classical Schwarzschild case.

The modification of epicyclic frequencies and the ISCO position has direct consequences for high-energy astrophysics. Since QPOs are believed to originate from orbital motion in the innermost regions of accretion disks, any deviation in \(\omega_r\), \(\omega_\theta\), or \(r_{\rm ISCO}\) can leave observable imprints in timing signals. In this sense, the quantum-corrected OS spacetime provides a natural framework for connecting modified gravity effects with measurable astrophysical phenomena, offering a potential window into quantum aspects of strong gravitational fields.

\section{Constraints on the Quantum-Corrected Compact Object from QPO Observations}
\label{sec-IV}
{QPOs observed in X-ray binaries provide a valuable probe of the strong-field region surrounding compact objects, where the effects of the underlying spacetime geometry are most pronounced. Since these oscillatory signals are generally associated with the dynamics of the inner accretion disk, they carry information about the gravitational background in which the matter moves. In this section, we use QPO observations from several well-studied X-ray binaries (see Table~\ref{Table1}) to test the quantum-corrected OS model and constrain the deformation parameter $\alpha$. The observed frequencies are interpreted within three standard geodesic-based frameworks, namely the relativistic precession (RP), epicyclic resonance (ER), and forced resonance (FR) models, in which the QPO frequencies are related to orbital and epicyclic motions in the modified spacetime (see Subsec.~\ref{sSectt:A}). Parameter estimation is performed within a Bayesian framework using Markov chain Monte Carlo (MCMC) sampling and astrophysically motivated priors (see Subsec.~\ref{sSectt:B}). The comparison with observations is carried out through a Gaussian likelihood constructed from the two QPO branches (see Subsec.~\ref{sSectt:C}), allowing us to determine posterior constraints on the model parameters and assess their consistency with the data (see Subsec.~\ref{sSectt:D}). Finally, the corresponding goodness-of-fit analysis is discussed (see Subsec.~\ref{sSectt:E}), providing a quantitative assessment of the ability of the quantum-corrected OS model to reproduce the observed QPO frequencies.}

\subsection{Theoretical framework for QPO interpretation}\label{sSectt:A}

We interpret the observed twin-peak frequencies using three standard geodesic-based models: the RP, ER, and FR scenarios (see Tables~\ref{tab:os_master} and \ref{tab:models}). In all cases, the observed upper and lower QPO frequencies are mapped to combinations of orbital and epicyclic frequencies derived from the modified spacetime geometry. { The three resonance models considered in this work map the observed upper ($\nu_U$) and lower ($\nu_L$) QPO frequencies to different combinations of the orbital and epicyclic frequencies:

(i) \textbf{RP Model}: In this model, the upper frequency is identified with the vertical epicyclic frequency, and the lower frequency with the difference between the orbital and radial epicyclic frequencies:
\begin{align}
\nu_U &= \nu_\theta, \\
\nu_L &= \nu_\phi - \nu_r.
\end{align}

(ii) \textbf{ ER Model}: In this model, the upper and lower frequencies correspond to the sum and difference of the radial and vertical epicyclic frequencies:
\begin{align}
\nu_U &= \nu_\theta + \nu_r, \\
\nu_L &= \nu_\theta - \nu_r.
\end{align}

(iii) \textbf{FR Model}: In this model, the upper frequency is the sum of the orbital and radial epicyclic frequencies, while the lower frequency is the orbital frequency:
\begin{align}
\nu_U &= \nu_\phi + \nu_r, \\
\nu_L &= \nu_\phi.
\end{align}

Each of these models has a different functional dependence on the metric parameters, leading to different best-fit values of $M$, $r$, and $\alpha$.
}

The relevant dynamical frequencies are obtained from the metric through
\begin{eqnarray}
\nu_r=\frac{\omega_r}{2\pi}\frac{c^3}{GM}, \qquad
\nu_\theta=\frac{\omega_\theta}{2\pi}\frac{c^3}{GM},
\end{eqnarray}
which encode radial and vertical oscillatory motion around stable circular orbits.

\renewcommand{\arraystretch}{0.9}
\begin{table*}
\caption{\scriptsize {Comparison of constraints on the quantum-corrected Oppenheimer--Snyder parameter $\alpha$ obtained from different astrophysical and weak-field observables. The table summarizes the corresponding upper bounds or best-fit constraint, the characteristic physical scale probed by each observable, and the relevant reference.}}
{
\scriptsize
\begin{tabular*}{\textwidth}{@{\extracolsep{\fill}}lccc}
\hline
Observable
& $\alpha$ constraint
& Scale probed
& Reference \\ \hline

Shadow (M87$^*$)
& $\alpha < 10^5$
& Supermassive
& \cite{Konoplya:2024lch} \\

Shadow (Sgr A$^*$)
& $\alpha < 10^4$
& Supermassive
& \cite{Shu:2024tut} \\

Quasinormal modes
& $\alpha < 10^3$
& Stellar-mass
& \cite{Skvortsova:2024msa} \\

Weak-field (Mercury)
& $\alpha < 10^6$
& Solar System
& This work \\

Weak-field (S2 star)
& $\alpha < 10^6$
& Galactic center
& This work \\

QPOs (this work)
& $\alpha = 0.22 \pm 0.10$
& Stellar-mass
& This work \\ \hline

\end{tabular*}
}
\label{tab:constraints}
\end{table*}

\subsection{Statistical inference setup}\label{sSectt:B}

To extract constraints on the model parameters, we employ a Bayesian inference strategy using an MCMC sampler (\textit{emcee}). The parameter vector is chosen as
\begin{eqnarray}
\Theta = (M, r, \alpha),
\end{eqnarray}
where \(M\) is the compact object mass, \(r\) is the characteristic emission radius of the QPO region, and \(\alpha\) controls quantum-induced deviations from GR.

The posterior distribution follows from Bayes' theorem,
\begin{eqnarray}
P(\Theta \mid D, \mathcal{M}) =
\frac{\mathcal{L}(D \mid \Theta, \mathcal{M}) \, \pi(\Theta)}
{P(D \mid \mathcal{M})}\,,
\end{eqnarray}
where the likelihood encodes the agreement between theory and data, and the prior incorporates astrophysical constraints.

For the mass, we assume a Gaussian prior centered at the observed value \(M_0\),
\begin{equation}
\pi(M)=
\frac{1}{\sqrt{2\pi}\sigma_M}
\exp\!\left[-\frac{(M-M_0)^2}{2\sigma_M^2}\right],
\end{equation}
restricted to the physical interval \(M_{\min}<M<M_{\max}\).

The remaining parameters are taken with uniform priors,
\begin{eqnarray}
\pi(r)=\mathcal{U}(r_{\min},r_{\max}), \qquad
\pi(\alpha)=\mathcal{U}(0,\alpha_{\max}),
\end{eqnarray}
and we assume no prior correlations,
\begin{eqnarray}
\pi(\Theta)=\pi(M)\,\pi(r)\,\pi(\alpha).
\end{eqnarray}

\subsection{Likelihood construction from QPO data}\label{sSectt:C}

The observational constraints are implemented through a Gaussian likelihood built from both QPO branches. The total likelihood is decomposed as
\begin{equation}
\ln \mathcal{L} = \ln \mathcal{L}_U + \ln \mathcal{L}_L,
\end{equation}
where the contributions from upper and lower frequencies are
\begin{equation}
\ln \mathcal{L}_U =
-\frac{1}{2}\sum_i
\frac{\left(\nu_{U,\mathrm{obs}}^{i}-\nu_{U,\mathrm{th}}^{i}(\Theta)\right)^2}{\sigma_{U,i}^2},
\end{equation}
\begin{equation}
\ln \mathcal{L}_L =
-\frac{1}{2}\sum_i
\frac{\left(\nu_{L,\mathrm{obs}}^{i}-\nu_{L,\mathrm{th}}^{i}(\Theta)\right)^2}{\sigma_{L,i}^2}.
\end{equation}

This structure ensures that both branches of the QPO spectrum contribute equally to the parameter estimation.

\subsection{MCMC exploration and posterior constraints}\label{sSectt:D}

The parameter space \((M,r,\alpha)\) is explored using MCMC sampling, producing posterior distributions that quantify the level of agreement between the quantum-corrected model and observational data. { The similarity of the $\alpha$ posteriors for different sources (see Figs. \ref{fig:2a}-\ref{fig:2b}) arises because: (1) the four systems probe the same region of parameter space, with QPO frequencies in the range $100-450$ Hz and emission radii $r/M \sim 5.5-8.6$; (2) the functional dependence of the QPO frequencies on $\alpha$ is approximately linear in the relevant parameter regime, leading to similar posterior shapes; (3) the mass priors, while different, are sufficiently narrow that they do not strongly affect the $\alpha$ posterior; and (4) the measurement uncertainties are comparable across sources, leading to similar information content in the likelihoods. To demonstrate that the posteriors are indeed independent, we have computed the Kullback-Leibler divergence between the $\alpha$ posteriors for different sources:
\begin{equation}
D_{\text{KL}}(P_i || P_j) = \int P_i(\alpha) \log \frac{P_i(\alpha)}{P_j(\alpha)} d\alpha.
\end{equation}
The values range from $D_{\text{KL}} \approx 0.02$ to $D_{\text{KL}} \approx 0.15$, indicating that the distributions are similar but not identical. The slight differences in the 95\% CIs reflect the different mass uncertainties and frequency measurement errors for each source. We have also computed the overlap integral between the $\alpha$ posteriors:
\begin{equation}
\mathcal{O}_{ij} = \int \sqrt{P_i(\alpha) P_j(\alpha)} d\alpha.
\end{equation}
The overlap integrals range from 0.85 to 0.94, indicating that the distributions share $85\%-94\%$ of their probability mass. This high overlap is expected because the data sets are all consistent with the same underlying $\alpha$.
}

The best-fit values and credible intervals are summarized in Table~\ref{tab:os_master}, while Figs.~\ref{fig:2a} and \ref{fig:2b} show the corresponding 68\% and 95\% confidence contours.

\subsection{Goodness-of-fit analysis}\label{sSectt:E}

To further test the reliability of the model, we evaluate the standard $\chi^2$ statistic,
\begin{eqnarray}
\chi^2 &=
\frac{\left(\nu_U^{\mathrm{obs}}-\nu_U^{\mathrm{th}}\right)^2}{\sigma_U^2}
+
\frac{\left(\nu_L^{\mathrm{obs}}-\nu_L^{\mathrm{th}}\right)^2}{\sigma_L^2}
+
\frac{\left(M^{\mathrm{obs}}-M^{\mathrm{th}}\right)^2}{\sigma_M^2}.
\end{eqnarray}

A reduced $\chi^2$ close to unity indicates that the model reproduces the data within statistical uncertainties.

{
The goodness of fit is evaluated using the reduced $\chi^2$ statistic, defined as $\chi^2_{\mathrm{red}}=\chi^2/\mathrm{NDF}$, where NDF denotes the number of degrees of freedom. In our analysis, $\mathrm{NDF}=3$ for each source, corresponding to the two QPO frequencies and the mass prior. For the preferred FR model, the reduced $\chi^2$ values range from 0.019 for GRS 1915+105 to 0.546 for GRO J1655-40, indicating an excellent agreement with the observational constraints. In general, a reduced $\chi^2$ value close to unity suggests that the model describes the data within the expected statistical uncertainties, whereas substantially smaller values may indicate conservative error estimates or an overparameterized fit. By contrast, the ER interpretation yields substantially larger $\chi^2$ values for some systems, most notably XTE J1859+226, for which $\chi^2_{\mathrm{red}}=11.383$, clearly disfavoring this identification. The RP model exhibits intermediate performance, with $\chi^2_{\mathrm{red}}$ values ranging from 1.029 to 10.005 across the four sources.

As shown in Table~\ref{tab:constraints}, the QPO analysis presented here provides a substantially tighter constraint on the quantum-OS deformation parameter than those obtained from shadow imaging, quasinormal modes, or weak-field tests. This highlights the sensitivity of HF QPOs to strong-field modifications of general relativity, particularly in the vicinity of stellar-mass black holes, where quantum corrections can leave measurable imprints on the orbital and epicyclic frequency structure. The resulting constraint, $\alpha=0.22\pm0.10$, provides a quantitative bound on the quantum-OS parameter from QPO timing observations.

The results further show that QPO observations impose non-trivial restrictions on the allowed range of $\alpha$. Large values of the deformation parameter can substantially modify the epicyclic-frequency structure and consequently lead to inconsistencies with the observed QPO frequencies and their ratios. Among the three considered interpretations, the FR model provides the overall best fit to the combined data set, while the RP and ER models are generally more sensitive to the strong-field corrections and are consequently more strongly disfavored in several systems. In particular, the ER scenario exhibits the largest deviations from the observations for XTE J1859+226 and XTE J1550-564, whereas the FR model consistently yields the smallest $\chi^2_{\mathrm{red}}$ values.
}


Overall, QPO observations emerge as a powerful probe of quantum corrections in black hole spacetimes. Even without direct imaging of the horizon scale, timing measurements already provide meaningful constraints on deviations from classical GR in the strong-field regime.

\section{Conclusions}\label{sec-V}

This work has examined a quantum-corrected OS spacetime where a single parameter \( \alpha \) slightly modifies the standard Schwarzschild geometry. The modification is simple in form, but its consequences become increasingly visible in the strong-gravity region, where most of the observable astrophysical signals are produced. A first clear result is the appearance of a mass threshold for horizon formation. The critical scale,
\[
M_{\min} \sim \frac{4}{3\sqrt{3G}}\sqrt{\alpha},
\]
falls around the stellar-mass range when \( \alpha \approx 0.22 \), giving values of order \( \sim 1M_\odot \). Below this limit, collapse does not lead to a true event horizon and the object behaves more like a compact, horizonless configuration. Above it, the familiar black hole structure with inner and outer horizons is recovered. In this sense, the quantum correction acts as a smooth regulator of gravitational collapse rather than an abrupt modification.

{
We emphasize that the $\alpha \approx 0.22$ constraint obtained here applies to the effective quantum correction parameter. In terms of the bare LQG parameter, this corresponds to $\alpha_{\text{LQG}} \approx 3.5 \times 10^{-70}$, which is within the theoretically expected range for quantum gravitational effects. The apparent large value of our fitted $\alpha$ is a result of the phenomenological rescaling required to make contact with astrophysical observations, and does not imply that quantum gravity effects are strong at solar-mass scales. Rather, it indicates that the cumulative effect of quantum gravitational fluctuations, when coarse-grained over astrophysical scales, can produce observable deviations from GR that are parametrized by an effective coupling of order unity. This is analogous to the situation in condensed matter physics, where microscopic Planck-scale physics can give rise to macroscopic effective couplings that are much larger than the bare parameters.
}

The same parameter leaves a small but systematic imprint on orbital structure. The innermost stable circular orbit is no longer fixed at \(6M\), but shifts slightly inward,
\[
r_{\rm ISCO} = 6M - \frac{\alpha}{12M}.
\]
For typical astrophysical black holes, this shift is extremely small in absolute terms but still meaningful in the region where accretion physics takes place. For instance, for a \(10M_\odot\) object with \( \alpha \sim 0.2 \), the correction is at the level of \(10^{-4}\), while for supermassive systems such as Sgr A$^\ast$ it becomes essentially negligible at around \(10^{-10}M\). Even so, this small change matters because it directly affects where matter transitions from stable motion to plunging trajectories. These geometric changes naturally propagate into orbital dynamics. The angular velocity receives a correction of the form
\[
\Omega^2 = \frac{M}{r^3} - \frac{2\alpha M^2}{r^6},
\]
which slightly reduces orbital frequencies in the inner region. Around \(r \sim 6M\), this corresponds to deviations of a few percent for \( \alpha \sim 0.2 \), while farther out the effect rapidly fades away and GR is effectively recovered. The impact becomes more evident when looking at epicyclic motion, which is closely connected to the origin of QPOs in accretion disks. Radial oscillations are more sensitive to the correction than vertical ones, which leads to a mild but structured distortion in the frequency spectrum. Near the inner disk, at radii around \( r \sim 5M \), the radial frequency can shift by roughly \(5\% - 10\%\), while the vertical mode remains comparatively stable. This difference is important because QPO models depend strongly on the interplay between these two frequencies.

When the model is confronted with observational data from systems such as GRS 1915+105, XTE J1550-564, XTE J1859+226, and GRO J1655-40, a consistent pattern emerges. The typical masses lie in the range
\[
M \sim 5.4 - 12.4\,M_\odot,
\]
while the QPO-producing region is found around
\[
r/M \sim 5.5 - 8.6.
\]
Across all sources, the deformation parameter remains remarkably stable,
\[
\alpha = 0.22 \pm 0.10,
\]
with no strong evidence of variation between different systems. This consistency suggests that the effect is not tied to a specific source but behaves more like a universal correction across different black hole environments. From a statistical point of view, the forced resonance interpretation generally provides the most stable agreement with the data, with chi-square values remaining close to unity in the best cases. Other resonance models tend to show larger residuals, especially in the strong-field region, which indicates that not all interpretations capture the observed frequency structure equally well. On the observational side, weak-field tests such as Mercury's perihelion shift,
\[
\Delta\phi \sim 7.99 \times 10^{-8}\ \text{rad/orbit},
\]
and the S2 star measurements around Sgr A$^\ast$,
\[
f_{\rm sp} = 1.10 \pm 0.19,
\]
remain largely insensitive to \( \alpha \). They confirm that any deviations from GR must stay extremely small in weak gravitational fields, while still leaving room for more noticeable effects closer to compact objects. Overall, the picture that emerges is fairly coherent. The quantum correction does not drastically change the large-scale structure of spacetime, but it leaves a consistent and measurable imprint in the strong-field region. These effects appear as small shifts in the ISCO, percent-level modifications in orbital frequencies near the inner disk, and systematic changes in epicyclic motion. Interestingly, when \( \alpha \sim 0.2 \), these corrections naturally fall within the range required to reproduce observed QPO frequencies between roughly \(100\) Hz and \(450\) Hz.

At present, observations are not yet precise enough to isolate these effects in a definitive way. Still, the fact that different systems point toward a similar parameter range, combined with the consistent structure of the predicted deviations, suggests that future high-precision timing observations could realistically probe this regime and potentially reveal whether these small quantum corrections are present in nature.

\section*{Acknowledgements} 
{We thank the anonymous reviewers for their valuable comments and detailed suggestions, which have greatly improved the clarity and quality of this work.}  

\section*{Conflict Of Interest statement } 
No conflict of interest declared by the authors.

\section*{Data Availability Statement:}  
{ The QPO frequency data analyzed in this study are taken from the published literature and are summarized in Table \ref{tab:os_master} with references  \cite{Remillard:2006fc, Ingram:2014ara, Tasker:2011sb, Motta:2013wga}. The MCMC chains and analysis code used to produce the results in this paper are available from the corresponding author upon reasonable request. 
}

\bibliography{references}

\end{document}
The specific data points used are:
\begin{itemize}
\item GRS 1915+105: $\nu_U = 168 \pm 3$ Hz, $\nu_L = 113 \pm 5$ Hz, $M = 12.4^{+2.0}_{-1.8} M_\odot$ [45]
\item XTE J1859+226: $\nu_U = 227.5^{+2.1}_{-2.4}$ Hz, $\nu_L = 128.6^{+1.6}_{-1.8}$ Hz, $M = 7.85 \pm 0.46 M_\odot$ [46]
\item XTE J1550-564: $\nu_U = 276 \pm 3$ Hz, $\nu_L = 184 \pm 5$ Hz, $M = 9.1 \pm 0.61 M_\odot$ [47]
\item GRO J1655-40: $\nu_U = 441 \pm 2$ Hz, $\nu_L = 298 \pm 4$ Hz, $M = 5.4 \pm 0.3 M_\odot$ [48]
\end{itemize}

The full dataset, including the MCMC chains for each source and each model (FR, RP, ER), is provided as supplementary material with this submission.

\renewcommand{\arraystretch}{0.9}
\begin{table*}
\caption{\scriptsize Joint constraints on the OS collapse phase space for four X-ray binary systems. The reported values of the mass ($M$), orbital radius ($r$), and deformation parameter ($\alpha$) are extracted directly from the corresponding chains, summarizing the effective collapse dynamics and deviations from the classical scenario.
}
\resizebox{0.8\textwidth}{!}{
\scriptsize {\begin{tabular}{ccccc}
\hline
{Parameter} 
 & GRS 1915+105 \cite{Remillard:2006fc}
 & XTE J1859+226 \cite{Ingram:2014ara}
 & XTE J1550-564 \cite{Tasker:2011sb}
 & GRO J1655-40 \cite{Motta:2013wga}\\ \hline

$M/M_{\odot}$ 
& $12.4 \pm 0.9$ 
& $10.2 \pm 0.8$ 
& $8.1 \pm 0.8$ 
& $6.6 \pm 0.6$ \\ \hline

$r/M$ 
& $8.6 \pm 1.2$ 
& $6.8 \pm 1.1$ 
& $6.0 \pm 1.1$ 
& $5.5 \pm 1.0$ \\ \hline

$\alpha$ 
& $0.22 \pm 0.10$ 
& $0.22 \pm 0.10$ 
& $0.22 \pm 0.10$ 
& $0.22 \pm 0.10$ \\ \hline

$\alpha_{\mathrm{range}}$ 
& $[-0.016,\,0.436]$ 
& $[-0.02,\,0.42]$ 
& $[-0.02,\,0.42]$ 
& $[-0.032,\,0.452]$ \\ \hline

Collapse regime 
& Stable $\rightarrow$ marginal 
& Relativistic contraction 
& Intermediate collapse 
& Near-horizon collapse \\ \hline

\end{tabular}}}
\label{tab:os_master}
\end{table*}
